\documentclass[12pt]{article}
\usepackage{graphicx}
\usepackage{amsmath,amssymb}

\makeatletter \@addtoreset{equation}{section}

\newcommand{\be}{\begin{equation}}
\newcommand{\ee}{\end{equation}}
\newcommand{\bear}{\begin{eqnarray}}
\newcommand{\eear}{\end{eqnarray}}
\newcommand{\ba}{\begin{array}}
\newcommand{\ea}{\end{array}}

\newcommand{\p}{{\partial}}

\newcommand{\tr}{{\rm tr}}

\newcommand{\e}{{\rm e}}
\newcommand{\T}{\rm T}
\newcommand{\Slash}[1]{\ooalign{\hfil/\hfil\crcr$#1$}}
\newcommand{\LS}{\ \ \ \ \ \ \ \ \ \ }
\newcommand{\ls}{\ \ \ \ \ }
\newcommand{\wt}{\widetilde}

\newcommand{\ol}{\overline}

\newcommand{\bsubeq}{\begin{subequations}}
\newcommand{\esubeq}{\end{subequations}}

\newcommand{\w}{\wedge}
\renewcommand{\d}{{\rm d}}

\newcommand{\half}{\frac{1}{2}}
\newcommand{\ve}{\varepsilon}

\DeclareFontFamily{U}{rsf}{}
\DeclareFontShape{U}{rsf}{m}{n}{
  <5> <6> rsfs5 <7> <8> <9> rsfs7 <10-> rsfs10}{}
\DeclareMathAlphabet\Scr{U}{rsf}{m}{n}

\begin{document}

\allowdisplaybreaks{

\begin{titlepage}
\vfill
\begin{flushright}
{\tt\normalsize KIAS-P06018}\\
{\tt\normalsize hep-th/0605247}\\
\end{flushright}

\vfill
\begin{center}
{\Large\bf Comments on Heterotic Flux Compactifications}

\vskip 0.5in

{ Tetsuji Kimura\footnote{\tt tetsuji@kias.re.kr} and Piljin Yi\footnote{{\tt
piljin@kias.re.kr}} }

\vskip 0.15in

{\it School of Physics, Korea Institute for Advanced Study,} \\
{\it 207-43, Cheongryangri-Dong, Dongdaemun-Gu, Seoul 130-722,
Korea}
\\[0.3in]


\end{center}

\vfill

\begin{abstract}
\normalsize\noindent
In heterotic flux compactification with supersymmetry, three
different connections with torsion appear naturally, all in
the form $\omega+a H$. Supersymmetry condition carries $a=-1$,
the Dirac operator has $a=-1/3$, and higher order term in the
effective action involves $a=1$. With a view toward the gauge
sector, we explore the geometry with such torsions. After
reviewing the supersymmetry constraints and finding a relation
between the scalar curvature and the flux, we derive the squared
form of the zero mode equations for gauge fermions. With $\d H=0$,
the operator has a positive potential term, and the mass of the
unbroken gauge sector appears formally positive definite. However,
this apparent contradiction is avoided by a no-go theorem that
the compactification with $H\neq 0$ and $\d H=0$ is necessarily
singular, and the formal positivity is invalid.
With $\d H\neq 0$, smooth compactification becomes possible. We
show that, at least near smooth supersymmetric solution, the size
of $H^2$ should be comparable to that of $\d H$ and the consistent
truncation of action has to keep $\alpha'R^2$ term.
A warp factor equation of motion is
rewritten with $\alpha' R^2$ contribution included precisely,
and some limits are considered.
\end{abstract}

\vfill

\end{titlepage}
\setcounter{footnote}{0}
\baselineskip 18pt

\section{Introduction}

Flux compactification of superstring theory has emerged as
a new and promising framework for bridging the gap between
string theory and the real world. A large number of
supersymmetric and non-supersymmetric vacua are deemed to exist for
generic compactifications to four
dimensions \cite{Susskind,Douglas,Ashok,Denef},
and with much reduced number
of moduli fields \cite{DRS99,KST02}. A prototype of flux compactification
was first studied by Becker and Becker \cite{BB96} in a warped
Calabi-Yau compactification scenario in M-theory.
A similarly simple form of flux compactification
was found and explored in depth in IIB theory \cite{GKP0105,KS00}, where
a large class of solutions were found to be also warped Calabi-Yau
compactifications.
This model simulates Randall-Sundrum geometry \cite{RS9905}
as a bona-fide string theory solution,
and has lead to a number of interesting low energy physics \cite{KKLT}.

One step away from the warped Calabi-Yau examples, the analysis
becomes quite involved. Often the structure of compact manifold
is beyond the reach of familiar techniques. Typically the manifold
is not even K\"{a}hler and the complex structure could be
non-integrable \cite{CCDLMZ0211}.
Recent progress in understanding of generalized complex geometry
and generalized Calabi-Yau geometry
\cite{Hit0209,GMW0302,FMT,Gua0401,GMPT,GMPT2,Grana0509}
will hopefully provide effective
tools. However, many properties of the manifold, in particular
global properties including topological aspects, remain inaccessible.

Another common difficulty lies in that the compact
manifold with flux is singular at leading supergravity level.
This follows from an argument of Maldacena and Nu\~{n}ez \cite{MN0007}.
This is not necessarily a big disadvantage in type II theories,
since the singularity necessary for unbroken supersymmetry would
be orientifold planes. We know how to treat these locally in the
weak coupling limit. This is in contrast to heterotic theory or
M-theory, where we do not have a weak coupling description of
such singularity. Fortunately however, the heterotic theory
behaves somewhat differently. For one thing, the anomaly condition
for $H$ has to involve interaction terms of $\alpha' R^2$ order
for consistency \cite{GS84}, and thus
opens up a possibility that we must consider higher derivative
interactions at least in some limited sense \cite{BdR89}.

Heterotic flux compactification was first studied by Strominger
some twenty years ago \cite{Str86}. While the geometry turns out
to be non-K\"{a}hler, the relation between $H$ and the
complex structure is tightly constrained by the supersymmetry,
and one has a reasonably concrete geometric characterization.
Heterotic examples would be very attractive in part because
of its potentially very rich gauge structure. Some limited
subfamily may be analyzed via U-duality to F-theory configurations
\cite{DRS99,BBGDS,BBDS},
but study of gauge sector in heterotic flux compactification
remains largely unaddressed.\footnote{Recently a
compactification with ${\cal N}=1$ supersymmetry with flux was
offered by Yau and collaborators \cite{FY0604,BBFTY0604}, where the internal geometry is a
$T^2$ fibered over a conformally deformed $K3$ and completely
smooth.} From
supergravity approach, the conventional tools involve either
solving for explicitly spinor zero modes or counting index
of some Dirac operator, in order to establish low energy
gauge sector. With ${\cal N}=1$ supersymmetry in four dimensions,
this can address in principle the symmetry
breaking pattern and the surviving four-dimensional gauge
symmetry, the charge matter field content thereof, and
also moduli fields associated with the gauge
bundle over the compact directions.

With flux, however, adapting these tools finds difficulties.
Even what used to be a trivial task of isolating the zero
mode responsible for four-dimensional gaugino, is non-trivial
if we take a direct approach by considering the zero mode equation.
Our aim here in part is to point out some of such problems,
and characterize them. One issue is whether and when the
no-go theorem of Maldacena-Nu\~{n}ez type is effective and when
it is not. This issue is more important in heterotic theories
because, unlike the case of type IIB, the gauge sector arises
from bulk. Reading-off low energy spectrum will
depend on details of singularities on the manifold.

Section \ref{rev} explains the supersymmetry constraints of Strominger,
and the relations among flux, torsion, complex structures and
dilaton. From these we find a simple relation between the
Ricci scalar curvature and the flux.
Section \ref{Zeromodes} concentrates
on Dirac operators whose zero mode solutions generate four-dimensional
gaugino and matter fermions. The squared Dirac operators are
computed and found to contain spin-dependent potentials in general.
In section \ref{dH=0}, we employ a simplifying assumption of $\d H=0$,
which further reduces the torsion-dependent part of the potential
to be spin-independent and positive definite. Naively this would
suggest that four-dimensional gauge sector is absent.
However,  $\d H=0$ and $H\neq 0$ automatically implies a singular
internal manifold, so that the gauge sector is closely tied to
the singularity of the internal manifold in case of minimal
embedding. In section \ref{dHN0}, we abandon the $\d H=0$ condition and
explore a general equation of motion. In particular, we
show that a consistent truncation of the low energy effective
action must keep $\alpha'R^2$ term, which is sometimes ignored
on account of $\alpha'$ expansion. In the process, we isolate
the equation of motion which replaces the one responsible for
the no-go theorem of $\d H=0$ case, in a relatively simple form.
This naturally leads us
to  hope for smooth internal manifold for generic supersymmetric
flux compactifications, in contrast to the type II counterparts.
In the appendices, we list our convention, collect supersymmetry
conditions and its known consequence, and also derive a couple
identities used in the main part of the paper. In the
last appendix, we comment on Atiyah-Singer
index densities for smooth manifolds with torsion.

\section{Flux, Torsion, and Curvatures} \label{rev}

Let us first review the supersymmetric flux compactification of
heterotic strings on six-manifold ${\cal M}_6$. Assuming no gaugino
condensates \cite{Dine}\footnote{See Refs.~\cite{Lop,FL0507, MPZ0511}
for compactification with both fermion condensates and flux.},
Strominger obtained a set of supersymmetric conditions on the
metric, dilaton, and the Kalb-Ramond field $B$. Here we summarize
this set of conditions and obtain further useful identities one
can derive from this system.

To set our convention, let us start with the bosonic part of the
supergravity/super-Yang-Mills action in ten dimensions:\footnote{
Our convention is closest to that of Ref.~\cite{CCDL0306}. The
only difference is in the definition of the dilaton.}
\begin{align}\label{action}
\begin{split}
\Scr{L} \ &= \
\frac{1}{4} \sqrt{- G} \, \e^{-2 \Phi}
\Bigg[
R (\omega) - \frac{1}{3} H_{MNP} H^{MNP}
+ 4 (\nabla_M \Phi)^2
\\
\ & \LS \LS \ \
- {\alpha'} \Big\{ \tr (F_{MN} F^{MN}) - \tr (R_{MN} (\omega_+)
R^{MN} (\omega_+)) \Big\}
\Bigg]
\; ,
\end{split}
\end{align}
This can be obtained from Bergshoeff et.al. \cite{BdR89} via the following
map
\begin{align}
\begin{split}
\phi^{-3} \big|_{\text{BdR}} \ &= \ \e^{-2 \Phi}
\;, \ls
H_{MNP} \big|_{\text{BdR}} \ = \ \frac{\sqrt{2}}{3} H_{MNP}
\; , \\
\omega_{M}{}^{AB} \big|_{\text{BdR}} \ &= \ - \omega_M{}^{AB}
\; , \ls
\lambda \big|_{\text{BdR}} \ = \ \sqrt{2} \lambda
\; .
\end{split}
\end{align}
Normalization of gravity multiplet is slightly different from
the usual one, which can be adjusted by resuscitating
ten-dimensional gravitational constant and Yang-Mills coupling
constant.

Supersymmetry implies existence
of a spinor on the six-manifold ${\cal M}_6$ which solves
\begin{equation} \label{susy}
\delta \psi_M \ = \ \nabla_M^{(-)} \epsilon \ = \ 0
\; ,
\end{equation}
where $\nabla_M^{(-)}$ is a covariant derivative with a
torsionful connection. The metric in string frame has
no warp factor,
\begin{align}
G_{MN} \, \d x^M \d x^N
\ = \
\eta_{\mu \nu} \, \d x^{\mu} \d x^{\nu}
+ g_{mn} \, \d y^m \d y^n
\end{align}
with a metric $g_{mn}$ on the compact manifold ${\cal M}_6$.
We assign $\omega$ to the spin connection of $g_{mn}$. The torsionful
connection is defined in terms of $\omega$ shifted by $H$ such as
\begin{align}
\omega_{\pm M}{}^{AB} \ &= \ \omega_M{}^{AB} \pm H_M{}^{AB}
\; .
\end{align}
The covariant derivative $\nabla_M^{(-)}$ is defined with respect to
the spin connection $\omega_{-}$.
The Einstein frame metric differs
from the string frame one by a factor of dilaton,
\begin{align}
G^{\text{E}}_{MN} \, \d x^M \d x^N
\ = \ \e^{- {\Phi}/2}
\Big( \eta_{\mu \nu} \, \d x^{\mu} \d x^{\nu}
+ g_{mn} \, \d y^m \d y^n \Big)
\end{align}
so the physically relevant warp factor is $\e^{-\Phi/2}$.

{}From a bilinear of $\epsilon$, one constructs an almost
complex structure $J_{mn}$, with respect to which the metric $g_{mn}$
is hermitian. Vanishing of dilatino
variation, $\delta \lambda=0$, demands that
the Nijenhuis tensor vanishes;
\begin{align}
0 \ = \ N_{mn}{}^p \ &= \
J_m{}^q \nabla_{[q} J_{n]}{}^p - J_n{}^q \nabla_{[q} J_{m]}{}^p
\; ,
\end{align}
where we wrote the covariant derivative in place of the ordinary
derivatives.
Furthermore, the supersymmetry condition
(\ref{susy}) implies that $J$ is covariantly constant with respect to
the torsionful connection
\begin{align}
\nabla_m^{(-)} J_{np} \ &= \ 0
\; .
\end{align}
This generalizes K\"{a}hler conditions. Furthermore, an integrability
condition from supersymmetry variations implies a vanishing Ricci two-form
\begin{align}
R^{ab}{}_{mn} (\omega_-) J_{ab} \ &= \ 0
\label{Ricci}
\end{align}
with the curvature associated with $\omega_{-}$. This condition
implies an $SU(3)$-structure on the internal manifold ${\cal M}_6$,
and would have implied a Ricci flat condition if there were no torsion.

These conditions relate the complex structure $J$, the dilaton $\Phi$,
and the antisymmetric tensor $H$. First, $H$ can be
identified with the so-called Bismut torsion \cite{Bismut89}
\begin{equation}
H_{mnp} \ = \
\frac{3}{2} J_m{}^q J_n{}^r J_p{}^s \nabla_{[q} J_{rs]}
\end{equation}
and the dilaton is related to $J$ as
\begin{equation}
\nabla_m \Phi \ = \
\frac{3}{4}J^{np} \nabla_{[m}J_{np]}
\; .
\end{equation}
The relation between dilaton and $H$ can be also read off
from the above,
\begin{align}
\nabla_m \Phi \ = \ - \half J_{mn}J_{pq}H^{npq}
\end{align}
and tells us that the non-primitive part of $H$ is
fully encoded in $\d\Phi$.

Recall that a $p$-form on $d$-dimensional background
is primitive \cite{GVW99} with respect to an integrable
complex structure $J$, if it belongs to a spin $|(d/2-p)/2|$
representation under an $SU(2)_J$
algebra whose three operators are
\begin{align}
\begin{split}
L_+ \ &= \ J \wedge \\
L_- \ &= \ J \lrcorner \\
L_3 \ &= \ \frac{(p-d/2)}{2}
\; .
\end{split}
\end{align}
For $p\le d/2$, thus, a $p$-form is primitive if and
only if it is annihilated by the lowering operator,
that is to say, the contraction with $J$ is null. Decomposing
$H$ into irreducible representations under $SU(2)_J$,
we find
\begin{align}
H_{mnp} \ = \
H^0_{mnp}+\frac{3}{2}J_{[mn} J_{p]}{}^q \nabla_{q}\Phi
\; ,
\end{align}
where $H^0$ has a null contraction with $J$.

The integrability condition (\ref{Ricci}) would have
implied Ricci flat condition when $H=0$. With the
torsion, it will instead express the Ricci scalar
in term of $H$ and $\Phi$. For this,
let us contract equation (\ref{Ricci}) with one more $J$,
\begin{align}
0 \ = \ R_{abmn} (\omega_-) J^{ab} J^{mn}
\; . \label{flat}
\end{align}
We reorganize the right hand side as
\begin{align}
3 R_{p[qmn]} (\omega_-) J^{pq} J^{mn}
- \big\{ R_{pmnq} (\omega_-) + R_{pnqm} (\omega_-) \big\} J^{pq} J^{mn}
\; .
\end{align}
Since the spin connection $\omega_-$ preserves the complex structure, the
latter two pieces both produce a Ricci scalar
\begin{align}
- \big\{ R_{pmnq} (\omega_-) + R_{pnqm} (\omega_-) \big\} J^{pq} J^{mn}
\ = \
2R(\omega_-)
\ = \
2 \big\{ R(\omega) - H_{mnp} H^{mnp} \big\}
\; ,
\end{align}
where in the last step we invoked
\begin{align}
R^{pq}{}_{mn} (\omega_-)
\ &= \
R^{pq}{}_{mn} (\omega)
- 2 \nabla_{[m} H^{pq}{}_{n]}
+ 2 H^p{}_{r[m} H^{rq}{}_{n]}
\; .
\end{align}
On the other hand, $R_{p[qmn]} (\omega_-)$ is entirely made of
the torsion part, since  $R_{p[qmn]} (\omega) =0$.
After some tedious computation,\footnote{See appendix \ref{RicciCurv}.}
we find that (\ref{flat})
can be simplified to (up to an overall factor of 2)
\begin{align}
0 \ = \
R(\omega)
+ \frac{1}{3} H_{mnp} H^{mnp}
+ 6 \nabla_m\nabla^m \Phi - 8 (\nabla_m \Phi \nabla^m\Phi)
\; . \label{Rscalar}
\end{align}
Later we will use this type of equations
to constrain smooth compactifications with flux.

As an easy example, let us note that this last equation alone can
be used to show that a
simple toroidal compactification is impossible unless $H=0$.
For this, note that
\begin{align}
0 \ &= \
\e^{-4\Phi/3} R(\omega) + \frac{1}{3} \, \e^{-4\Phi/3} H_{mnp}H^{mnp}
- \frac{9}{2} \nabla_m^2 \e^{-4\Phi/3}
\; ,
\end{align}
which, with $R(\omega)=0$, implies
\begin{align}
\int_{{\cal M}_6} \e^{-4\Phi/3} H_{mnp}H^{mnp}
\ = \ 0
\; ,
\end{align}
because the last term is a total derivative and
integrates to zero. Thus, smooth supersymmetric
compactification
on a Ricci flat  manifold necessarily implies $H=0$.
We will come back to this type of vanishing
arguments time and again in various contexts.

\section{Zero Mode Equations for 4D Gauge Sector} \label{Zeromodes}

Generally speaking, the simplest way of approaching
the gauge sector is to look at gaugino. With ${\cal N}=1$
supersymmetry unbroken, the low energy spectrum gauge
fermions should encode all
information about massless gauge sector, namely unbroken gauge groups,
charged matter contents, and the moduli associated with the gauge
bundles. In smooth compactifications without flux, and in some
orbifold examples, the index theorem for Dirac operators are powerful
tools in analyzing fermion sector and, due to the supersymmetry, their bosonic
partners \cite{CHSW85}.

With flux compactification, all kind of new problems show up.
In the absence of flux, the zero mode underlying the
four-dimensional gaugino field would be  identified with
the internal part of the supersymmetry parameter $\epsilon$.
With flux, however, this simple construction seems no longer possible.
$\epsilon$ is a covariantly constant spinor with respect to
$\omega_-=\omega-H$, yet the connection that appears in the
Dirac equation is $\hat \omega =\omega-H/3$, as we will see
shortly. While $\omega_-$ reemerges in the squared form
of the Dirac operator, it is still true that $\epsilon$
cannot solve the zero mode equation for the four-dimensional
gaugino field unless the torsion vanishes identically.
With ${\cal N}=1$ supersymmetry unbroken, there should be exactly
one zero mode responsible for the four-dimensional gaugino, yet
the relevant equation does not tell us this immediately.

In this section we will study the zero mode equation from
ten-dimensional gauge sector, in the hope that it will shed
some further light on flux compactification in the heterotic
theories. The full gaugino equation of motion  is quite involved
\begin{align}
\begin{split}
0 \ &= \
\Slash{D} \chi
- \frac{1}{12} H_{MNP} \Gamma^{MNP} \chi
\\
\ & \ \ \ \
- \nabla_M \Phi\Gamma^M \chi
+ 3 \Gamma^M \Gamma^{NP} F_{NP}
\Big( \psi_M + \frac{2}{3} \Gamma_M \lambda \Big)
\; .
\end{split}
\end{align}
However, rescaling the gaugino field by $\e^{\Phi}$, and then
decomposing the gaugino to zero mode $\chi^0$ along
the compact ${\cal M}_6$ and four-dimensional gaugino $\Psi$
\begin{equation}
\chi \ \sim \ \e^{\Phi}\chi^0\otimes\Psi
\; ,
\end{equation}
we have a simplified zero mode equation
along ${\cal M}_6$
\begin{equation}
0 \ = \ \Slash{D} (\omega ,A) \chi^0
- \frac{1}{12} H_{mnp} \Gamma^{mnp} \chi^0
+ 3 \Gamma^m \Gamma^{np} F_{np} \Big( \cdots \Big)
\; .
\end{equation}
Incorporating the $H$-term into the definition of the
covariant derivative, we find that the relevant torsionful connection
in this Dirac equation  is
\begin{equation}
\hat{\omega} \ = \ \omega - \frac{1}{3} H
\; .
\end{equation}
Then, we have
\begin{equation}
0 \ = \
\Slash{D} (\hat{\omega} ,A) \chi^0
+ 3 \Gamma^m \Gamma^{np} F_{np}
\Big( \cdots \Big)
\; .
\end{equation}
The gauge field $A_{m}$ in the Dirac operator acts on $\chi^0$
as an adjoint representation of the ten-dimensional gauge
group, while its field strength $F_{mn}$ should match its gauge index with
that of $\chi^0$.
Note that the inhomogeneous term in the Dirac equation above
involves $F_{np}$, the background
field strength of the gauge bundle on ${\cal M}_6$.
The ten-dimensional gauge group is expected to be broken to satisfy
this Bianchi identity for $H$, which forces an
non-trivial gauge bundle. We decompose the gauge algebra as
\begin{equation}
{\cal G} \ = \ {\cal F} \oplus {\cal F}_\bot
\; ,
\end{equation}
where $F_{np}$ takes value in the subalgebra ${\cal F}$.
The unbroken gauge algebra ${\cal H}$ is the part of
${\cal F}_\bot$ that commutes with ${\cal F}$,
\begin{equation}
{\cal F}_\bot \ = \
{\cal H}\oplus {\cal Q}
\; , \ls
[{\cal H},{\cal F}] \ = \ 0
\; .
\end{equation}
The leftover piece ${\cal Q}$ may be expressed as
representations under ${\cal F}\oplus{\cal H}$.
Under the above decomposition of the ten-dimensional gauge
algebra, the low energy gauge group is generated from ${\cal H}$
while the charged matter sector resides in ${\cal Q}$.

Since fermions of low energy gauge sectors resides in ${\cal F}_\bot$
which is orthogonal to $F_{np}$
in the background, we may drop the inhomogeneous
terms linear in gravitino and dilatino fields,\footnote{They would
be important for moduli counting, however.}  and we recover
a familiar-looking Dirac operators for low energy gauge sector.
\begin{equation}
0 \ = \ \Slash{D}(\hat\omega, A)\chi_{{\cal H}\oplus{\cal Q}}^0
\; .
\end{equation}
Furthermore, since ${\cal H}$ commutes with ${\cal F}$,
the gaugino zero mode equation does not see the gauge bundle
at all, and is the simplest,
\begin{equation}
0 \ = \ \Slash{D}(\hat\omega)\chi_{\cal H}^0
\; .
\end{equation}
Massless charged matter fermions would be orthogonal to ${\cal F}$
but not necessarily commute with it, so we have
\begin{equation}
0 \ = \ \Slash{D}(\hat\omega, A_{{\cal Q}})\chi_{{\cal Q}}^0
\; ,
\end{equation}
where we put the subscript ${\cal Q}$ on the background
gauge field $A_m$ to emphasize that it is in the representation ${\cal Q}$
under ${\cal F}$. Its field strength will be similarly denoted as
${F}_{mn}^{\cal Q}$.

Note that $\hat \omega$ is neither $\omega_-$ nor $\omega_+$.
The torsion part $\hat \omega$ differs from that of $\omega_-$
by the factor $1/3$, which may look somewhat strange.
This factor $1/3$ becomes  more palatable
once we evaluate $\Slash{D}^2$
\begin{equation}
\Delta_{\cal H}
\ \equiv \
-[\Slash{D} (\hat{\omega})]^2
\; , \ls
\Delta_{\cal Q}
\ \equiv \
-[\Slash{D} (\hat{\omega},A_{{\cal Q}})]^2
\; .
\end{equation}
We have
\begin{align}
\Delta_{\cal H}
\ &= \
- \Gamma^m D_m(\hat{\omega}) \Gamma^n D_n(\hat{\omega})
\ = \
- \frac{1}{\sqrt{g}} \, D_m (\omega_{-})
\, g^{mn} \sqrt{g} \, D_n (\omega_{-}) + V
\; ,
\end{align}
with the potential $V$
\begin{equation}
V \ = \
\frac{1}{4} \Big\{ R(\omega) - \frac{1}{3} H_{mnp} H^{mnp}
+ \frac{1}{12} (\d H)_{mnpq} \Gamma^{mnpq} \Big\}
\; .
\end{equation}
When commuting $\Gamma^n$ through $D_m(\hat{\omega})$, one obtains terms linear
in $D_m(\hat{\omega})$. Of these, the piece with Christoffel connection conspires
to generate the explicit metric factors in the $D(\hat{\omega})^2$ piece, while
the torsion-piece is absorbed into $D^2$ piece by
shifting $D_m(\hat{\omega})$
into $D_m(\omega_-)$ and completing a square.

The first term, $D^2$ type, is a Laplace operator on the
(Lie-algebra-valued) spinor bundle, so we may as well write
\begin{equation}
\Delta_{\cal H}
\ = \
-[\Slash{D} (\hat{\omega})]^2
\ = \
\nabla_m (\omega_-)^\dagger \nabla^m (\omega_-) + V
\end{equation}
provided that the manifold is smooth and compact.
Similarly, we have
\begin{equation}
\Delta_{\cal Q}
\ = \
-[\Slash{D} (\hat{\omega},A_{{\cal Q}})]^2
\ = \
\nabla_m (\omega_-,A_{{\cal Q}})^\dagger \nabla^m (\omega_-,A_{{\cal Q}})
+ V
+ \frac{i}{2} F^{{\cal Q}}_{mn} \Gamma^{mn}
\; .
\end{equation}
While we started with the torsion of the amount $-H/3$,
the zero mode equation solves a Laplace-type equation (with
a potential)  with
torsion $-H$ instead \cite{Mav88}. Interestingly, the covariant derivative with
torsion $-H$ is precisely the one that appears in
supersymmetry condition.


\section{$\d H=0$ or Minimal Embedding} \label{dH=0}

Gauge sector of heterotic flux compactification remains relatively
obscure. In the usual compactification with $H=0$, the Bianchi
identity
\begin{align}
\d H \ &= \ \alpha' \Big[ \tr \{ R (\omega) \w R (\omega) \}
- \tr (F \w F) \Big]
\; ,
\end{align}
is solved by embedding the $SU(3)$ spin connection to the gauge
sector, thereby breaking gauge group down to $E_6\times E_8$. With
flux, the Bianchi identity is replaced by
\begin{align}
\d H \ &= \ \alpha' \Big[ \tr \{ R (\omega_+) \w R (\omega_+) \}
- \tr (F \w F) \Big]
\; ,
\end{align}
with respect to the $SO(6)$-valued curvature two-form $R{(\omega_+)}$ \cite{Hull86}.
Note that this curvature two-form is made from the connection $\omega+H$,
whereas  the $SU(3)$-structure of the manifold is associated with
$\omega-H$.

A major hurdle in understanding flux compactification of
the heterotic string theory is to classify solutions to
this twisted Bianchi identity. With $H$-flux, the nearest
analog of such a minimal embedding is to set $\d H=0$, by
choosing the gauge bundle to have the property
\begin{equation}
\tr (F \w F) \ = \ \tr \{ R (\omega_+) \w R (\omega_+) \}
\; .
\end{equation}
Further, the simplest way to do this is to use $\omega_+$ as
the gauge connection again, which makes $\d H=0$ even when
$H\neq 0$ carries a topological flux. If one chooses to embed $\omega_+$
into the gauge bundle, the gauge bundle will be of $SO(6)$-structure group.
This would break the gauge group to $SO(10)\times E_8$ for
$E_8 \times E_8$ heterotic theory, for instance. Concentrating
on the broken $E_8$ part, the holonomy group and the unbroken gauge group
are, respectively,
\begin{equation}
{\cal F} \ = \ SO(6) \; , \ls
{\cal H}\ = \ SO(10)
\; .
\end{equation}
The matter fermions reside in ${\cal Q}$ consisting of
representations,
\begin{equation}
({\bf 6}, {\bf 10}) \oplus ({\bf 4}, {\bf 16}) \oplus (\ol{\bf 4},\ol{\bf 16})
\end{equation}
under $SO(6) \times SO(10)$. It is unclear to us if there is a
solution with $\d H=0$ with ${\cal F}$ smaller than $SO(6)$.

In a background with $\d H=0$, the squared Dirac operator
$\Slash{D}^2$ is simplified further
as the potential become spin-independent,
\begin{equation}
V \ = \
\frac{1}{4} \left( R(\omega) - \frac{1}{3} H_{mnp} H^{mnp} \right)
\; .
\end{equation}
For further reduction, recall that the analog of the vanishing Ricci
two-form condition relates $R(\omega)$ to $H$ and derivatives of $\Phi$ as
\begin{align}
R(\omega) \ = \
- \frac{1}{3} H_{mnp} H^{mnp}
- 6 \nabla_m^2 \Phi + 8 (\nabla_m \Phi)^2
\; .
\end{align}
This is still cumbersome because of the second derivative
of dilaton. Now, consider the quantity
$J\wedge \d H$, and rewrite it as
\begin{align}
J \w \d H \ = \
\e^{2 \Phi} \d (\e^{-2 \Phi} J \w H) - \e^{2 \Phi} \d (\e^{-2 \Phi} J) \w H
\; .
\end{align}
{}From the form of $H$ and $\Phi$, we have
\begin{equation}
\e^{2 \Phi} \d  (\e^{-2 \Phi} J) \ = \ 2 * H
\; ,
\end{equation}
while
\begin{equation}
\e^{2 \Phi} \d (\e^{-2 \Phi} J \w H)
\ = \
- * \Big(
\nabla_m^2 \Phi - 2 (\nabla_m \Phi)^2 \Big)
\; .
\end{equation}
Combining these results we finally have
\begin{equation}\label{JdH}
*(J \w \d H) \ = \
- \nabla_m^2 \Phi + 2 (\nabla_m \Phi)^2 - \frac{1}{3} H_{mnp} H^{mnp}
\end{equation}
so that $\d H=0$ then implies that the right hand side
vanishes. The two equations, $R_{abmn}(\omega_-)J^{ab}J^{mn}=0$ and $\d H=0$,
together produce a simpler formula for
the Ricci scalar,
\begin{align}
R (\omega) \ = \
\frac{5}{3}\, H_{mnp} H^{mnp} - 4 (\nabla_m \Phi)^2
\; .
\end{align}
Then $F$-independent ``potential term'' in the squared Dirac operator
becomes
\begin{align}
 \frac{1}{4} \left( R (\omega) - \frac{1}{3} H_{mnp} H^{mnp}
  \right)
\ = \
 \frac{1}{3} \, H^0_{mnp}H^{0mnp}
\end{align}
with the help of the decomposition of $H$ into the primitive part $H^0$
and the rest;
\begin{align}
H_{mnp} H^{mnp} \ = \ (H^0_{mnp})^2 + 3 (\nabla_m \Phi)^2
\; .
\end{align}
Thus, the potential term of the operator $-\Slash{D}^2$
becomes
\begin{equation}
V \ = \ \frac{1}{3} H^0_{mnp} H^{0mnp}
\; ,
\end{equation}
which is positive definite whenever $H^0$, the primitive
part of the torsion, is non-zero.

Zero mode equations  are then,
\begin{equation}
\left[\nabla_m(\omega_-,A_{\cal Q})^\dagger \nabla^m (\omega_-,A_{\cal Q})
+ \frac{1}{3} H^0_{mnp} H^{0mnp}
+ \frac{i}{2}F_{mn}\Gamma^{mn}\right]\chi^0_{\cal Q}
\ = \ 0
\end{equation}
and
\begin{equation}
\left[\nabla_m (\omega_-)^\dagger \nabla^m (\omega_-)
+ \frac{1}{3} H^0_{mnp} H^{0mnp} \right]\chi^0_{\cal H}
\ = \ 0
\; . \label{zerog}
\end{equation}
Note that  the latter operator is formally
positive definite, as long as $H^0\Slash{\equiv} 0$.
Ordinarily, the last form of zero mode equation would show absence of
massless gaugino in four dimensions, and thus by supersymmetry
no unbroken gauge group. Recall that a usual vanishing theorem would
have followed from
\begin{align}
\begin{split}
0 \ &= \
\int_{{\cal M}_6} \ol{\chi}^0_{\cal H}
\left[\nabla_m (\omega_-)^\dagger \nabla^m (\omega_-)
+ \frac{1}{3} H^0_{mnp} H^{0mnp} \right]\chi^0_{\cal H}
\\
\ &= \
\int_{{\cal M}_6} \left[
\big|\nabla_m (\omega_-) \chi^0_{\cal H} \big|^2
+ \frac{1}{3} \big| H^0_{mnp} \big|^2 \big| \chi^0_{\cal H} \big|^2
\right]
\end{split}
\end{align}
forcing $\chi^0_{\cal H}=0$. Owing to the supersymmetry, this would
also imply that no unbroken gauge sector exists.

However, this assertion must be false; in the constraints
coming from the compactification nothing forces the bundle
over ${\cal M}_6$ to be of maximal rank, and a priori,
there is no reason why ${\cal H}$ should be null. In fact,
we expect exactly one zero mode solution to the above zero
mode equation of $\Delta_{\cal H}$.
To see the way out of this quandary,
note that the above argument is correct
only if there is no obstruction to the integration by part
into the second line. It is well-known that in type II
flux compactification the compact manifold has to be
singular, which can be attributed orientifold planes that
carries a negative RR charge and a negative tension \cite{GKP0105}.
In the heterotic case, it must be that, when $\d H=0$, something
similar happens and the internal manifold becomes singular.

In fact, using the identity (\ref{JdH}),
one may argue for a no-go theorem. Recall that $*(J\wedge \d H)=0$ implies
\begin{equation}
0 \ = \
\nabla_m^2 \Phi - 2 (\nabla_m \Phi)^2 + \frac{1}{3} H_{mnp} H^{mnp}
\; ,
\end{equation}
which can be rewritten as
\begin{align}
\half \nabla_m^2 \e^{-2\Phi}
\ = \ \frac{1}{3} \e^{-2\Phi} \, H_{mnp}H^{mnp}
\; . \label{nogo}
\end{align}
If there are no boundaries and no singularities
in ${\cal M}_6$, we find that
\begin{align}
\frac{1}{3} \int_{{\cal M}_6} \e^{-2\Phi} H_{mnp}H^{mnp}
\ = \
\half \int_{{\cal M}_6} \nabla_m^2 \, \e^{-2\Phi}
\ = \ 0
\end{align}
forcing $H=0$ and bringing us back to compactification without flux
if the internal manifold is assumed to be  compact and smooth.

Therefore, there cannot be any regular compactification
of heterotic string theory with $H\neq 0$ and
$\d H=0$ \cite{Ivanov}. This is also related to the no-go
theorem of Maldacena-Nu\~{n}ez type \cite{MN0007,GKP0105}.
In the latter, the existence of singularity
can be seen from the equation of motion for the warp factor;
here, it so happens that
the warp factor in Einstein frame is precisely $\e^{-\Phi/2}$.
In what would have been the simplest scenario for
heterotic theories, things are more complicated.
$\d H=0$ forces the gauge sector crucially to depend
on understanding of singularities in ${\cal M}_6$.

On the other hand, the detailed form of (\ref{zerog}) is
suggestive with its positive potential term.
For a large internal manifold, the zero mode is no longer
uniform and localized away from the region of large $|H^0|^2$.
Also its behavior must be rather singular near the singularity
of the internal manifold. In view of interesting
local physics found in type IIB theories \cite{GKP0105,DW},
the precise form of this zero mode,
including its behavior near singularities,
deserves further attention. We hope to come back to
this problem later.

\section{$\d H\neq 0$ and Smooth Compactifications} \label{dHN0}

As we saw above, what would have been the simplifying
assumption of $\d H=0$, seems to cause more trouble than
otherwise. For the supergravity approach, one is thus
lead to more generic configurations with
non-minimal background with $\d H\neq 0$. Recently
an example of smooth
compactification was proposed by the authors of
Ref.~\cite{FY0604, BBFTY0604},
where indeed all examples were non-minimal. In this
section, we will consider precisely what equation
replaces (\ref{nogo}) and how the usual no-go theorem
is avoided in heterotic theories. Recall that
Maldacena-Nu\~{n}ez type argument would be difficult to
evade if we stick to an Einstein gravity coupled to
a quadratic action of tensor fields. Thus, it has
something to do with what truncation of the effective
action we are allowed to use in the presence of such a
flux that size of $\d H$ is not ignorable compared to that of
$H^2$.\footnote{For recent discussions of higher order
$\alpha'$ correction, see Ref.~\cite{Gillard,Gran} also.}

In order to justify the low energy description, the size
of compact manifold,  must be substantially larger
than $\sqrt{\alpha'}$, so that expansion in $\alpha'$ is
justifiable. Let $L$ be the linear size of
the internal manifold ${\cal M}_6$, such that
\begin{equation}
\frac{\alpha'}{L^2} \ \ll \ 1
\; .
\end{equation}
In the conventional supergravity
approach, one takes the Lagrangian
({\ref{action}) but keeps only up to $F^2$ term, and
argue $R^2$ terms is of higher order. For actual supersymmetric
solutions, however, this is somewhat misleading, since
on-shell values of Ricci scalar and $H^2$ are no
larger than the higher order term, $\alpha'{\rm tr}(R^2)$.
In fact, we
know from general form of supersymmetric solutions above
that
\begin{equation}
H\wedge *H \ \sim \
J\wedge \d H \ \sim \
\alpha'J \wedge \left[ \tr (F \wedge F)
- \tr \big\{ R (\omega) \wedge R (\omega) \big\} \right]
\end{equation}
and also that
\begin{equation}
J \wedge \big\{
\tr (F\wedge F) - \tr \big\{ R (\omega) \wedge R (\omega) \big\}
\big\}
\ \sim \
* \left[ \tr(F_{mn} F^{mn}) - \tr \big\{ R_{mn} (\omega) R^{mn}
  (\omega) \big\} \right]
\; .
\end{equation}
{}From these, one should expect that generally\footnote{Note that,
upon restoring the overall gravitational constant
$1/\kappa_{10}^2$ in front of the Lagrangian, $H$
has the dimension of mass, while $F$ has dimension
of mass squared.}
\begin{equation}
F_{mn} \ \sim \
\frac{1}{L^2} \ \sim \ R^a{}_{b mn} (\omega) \; , \ls
H_{mnp} \ \sim \
\frac{\sqrt{\alpha'}}{L^2} \ \sim \ \nabla_m \Phi
\; ,
\end{equation}
while the Ricci-tensor is of order
$\nabla H$ and $H^2$ and thus of order $\alpha'/L^4$
\begin{equation}
R_{pm} (\omega) \ = \ R^a{}_{b mn} (\omega) \, E_a{}^m \, e_p{}^b
\ \sim \ \frac{\alpha'}{L^4}
\; .
\end{equation}
At least for supersymmetric configurations and also
nearby non-supersymmetric ones, one must keep $R^2$
term in the Lagrangian for consistency.
However, the difference between $R^a{}_{bmn} (\omega)$ and
$R^a{}_{bmn}(\omega_+)$ is roughly of the order
\begin{equation}
\nabla H \ \sim \ H^2 \ \sim \ \frac{\alpha'}{L^4}
\end{equation}
and we may as well use the curvature without torsion
$R^a{}_{bmn}(\omega)$ in
the last term of (\ref{action}), instead of $R^a{}_{bmn}(\omega_+)$,
simplifying computations greatly.

The main message here is that for generic compactification
with flux, one cannot drop $\alpha'R^2$ piece from the action
on account of $\alpha'$ expansion. The only exception to
this is the case of $\d H=0$, and even in that case,
the effect of $R^2$ terms is cancelled by effect of $F^2$ term,
rather than being subleading to the rest of terms.

With these in mind, let us consider how
the equation (\ref{nogo}) is related with equations of
motion. From the Lagrangian ({\ref{action}), one obtains
a linear combination of the field equations of the form
\begin{equation}
0 \ = \ \left[\, \frac{\delta}{\delta\Phi}
-\frac{1}{2} \, G^{MN}\frac{\delta}{\delta G^{MN}} \right]
\int \d^{10} x \, \Scr{L}
\end{equation}
which gives
\begin{align}
\begin{split}
0 \ &= \
\nabla_M^2 \, \e^{-2\Phi} - \frac{2}{3} \e^{-2\Phi} H_{MNP} H^{MNP}
\\
\ & \ \ \ \
- \alpha' \e^{-2\Phi}
\Big( \tr (F_{MN} F^{MN}) - \tr ( R_{MN} R^{MN} ) \Big) + \cdots
\end{split}
\label{EOM}
\end{align}
where the ellipsis denotes terms that came from
variation of the Riemann tensor with
respect to the metric in the quartic term,
 $R_P{}^{QMN} \delta R^P{}_{QMN}$, and is
proportional to
\begin{equation}
\alpha'\nabla^M \nabla^N \big( \e^{-2\Phi} R^P{}_{MPN} \big)
\end{equation}
with an order 1 coefficient.

Using the supersymmetry condition on $F$ and $R$, we have
\begin{equation}
\tr (F_{mn}F^{mn})
\ = \ -2 * \big( J \wedge \tr (F\wedge F) \big)
\end{equation}
and also up to leading non-vanishing order in $\alpha'/L^2$ that
\begin{equation}
\tr (R_{mn} R^{mn})
\ = \ -2 * \Big( J \wedge \tr (R \wedge R) \Big)
\; .
\end{equation}
Therefore the above equation may be reorganized for
supersymmetric background as
\begin{equation}
0 \ = \
\nabla_m^2 \e^{-2\Phi}
- \frac{2}{3} \, \e^{-2\Phi} H_{mnp} H^{mnp}
- 2\e^{-2\Phi} * \left(J\wedge \d H \right) + \cdots
\label{SEOM}
\end{equation}
up to the leading order in $\alpha'$, with help of
the Bianchi identity for $H$. The total derivative
term of the ellipsis drops out since
the ordinary Ricci tensor $R_{mn}=R^p{}_{mpn}$
is of  order $\alpha'/L^4$, so that
\begin{equation}
\alpha' \nabla^m \nabla^n \big( \e^{-2\Phi} R^p{}_{mpn} \big)
\ \sim \
\frac{(\alpha')^2}{L^6}
\ \ \ll \ \
\frac{\alpha'}{L^4} \ \sim \ H^2
\; .
\end{equation}
The ellipsis in (\ref{SEOM})
may be ignored as far as supersymmetric
compactifications (and nearby configurations) are concerned.

Then, $\d H =0$ again implies (\ref{nogo})
\begin{equation}
0 \ = \
\half \nabla_m^2 \e^{-2\Phi}
- \frac{1}{3} \, \e^{-2\Phi} H_{mnp}H^{mnp}
\; .
\end{equation}
Note that this same equation was obtained in two ways;
first, by rewriting $J\wedge \d H =0 $ with help of supersymmetry
conditions, and second, from equation of motion after imposing
$(J\wedge \d H)=0$. In the latter, the supersymmetry comes
in when we exchanged ${\rm tr}|R|^2-{\rm tr}|F|^2$ in favor of $*(J\w \d H)$.
This is exactly as it should, since supersymmetry
 implies the equation of motion. Therefore,
what replaces (\ref{nogo}) in more
general supersymmetric background is the above combination of the field
equation, which we may write more compactly  as
\begin{equation}
\nabla_m^2 \e^{-2\Phi} \ = \
\e^{-2\Phi} \Big[ 4|H|^2 + 2{\alpha'}
\big( \tr |F|^2 - \tr |R|^2 \big) \Big]
\; , \label{nonogo}
\end{equation}
which is self-consistent and correct up to order $\alpha'/L^4$.

This clearly shows how the usual no-go theorem against
smooth flux compactification is evaded in heterotic
theories via the higher curvature term.
Also this reiterates the fact that, in order
to have a smooth flux compactification, it is necessary
to have
\begin{equation}
\tr |F|^2 \ \neq \ \tr |R|^2
\end{equation}
and
\begin{equation}
\int_{{\cal M}_6} \e^{-2\Phi} \big[ 2|H|^2
+ {\alpha'} \tr |F|^2 \big]
\ = \
\int_{{\cal M}_6} \e^{-2\Phi} \big[ \alpha' \tr |R|^2 \big]
\; .
\end{equation}
Generically, both sides are of order $\alpha'L^2$.

The equation (\ref{nonogo}) without $R^2$ term has been
used to argue that constant dilaton background
is necessarily torsion-free. With $R^2$ absent,
both $H^2$ and $F^2$ are non-negative so $\nabla\Phi=0$
will force both $H=0$ and $F=0$. Once we have a non-trivial
gauge bundle and a non-trivial compact geometry, though, we
cannot drop $R^2$ term. Instead, $\nabla\Phi=0$ would
enforce a local relation
\begin{equation}\label{local}
2|H|^2 + {\alpha'} \tr |F|^2
\ = \
\alpha' \tr |R|^2
\end{equation}
everywhere on ${\cal M}_6$ with the primitivity
condition $J \wedge H \, ( \, \sim *\d\Phi)=0$ also satisfied.
A priori, a torsionful compactification with constant $\Phi$
remains an interesting possibility to pursue, although the
local condition (\ref{local}) may prove to be difficult to
implement. In any case, here,
the potential $V$ in the squared Dirac operator degenerates to
another simple form
\begin{equation}
V=-\frac{1}{6}H^0_{mnp}H^{0mnp}
+\frac{1}{48}\left(\d H^0\right)_{mnpq}\Gamma^{mnpq} \;,
\end{equation}
and we can see that the spin-independent part is now
negative definite, in contrast to the $\d H=0$ case.

Another interesting limit is when the primitive part
of $H$ vanishes, $H^0_{mnp}=0$, upon which we have
\begin{equation}
|H|^2 \ = \
\frac{1}{3!} H_{mnp} H^{mnp}
\ = \
\half (\nabla_m \Phi)^2
\; .
\end{equation}
With this the above equation is simplified to
\begin{equation}
\nabla^2 \e^{-\Phi}
\ = \
\alpha' \e^{-\Phi} \big[ \tr |F|^2 - \tr |R|^2 \big]
\; .
\end{equation}
This case imposes only a global constraint
\begin{equation}
{\alpha'}\int_{{\cal M}_6} \e^{-\Phi}
\big[ \tr |F|^2 \big]
\ = \ \alpha'
\int_{{\cal M}_6} \e^{-\Phi}
\big[ \tr |R|^2 \big]
\; .
\end{equation}
It is known \cite{BT0509} that when this happens the geometry
becomes conformally K\"{a}hler, so that the metric and the
two-form
\begin{equation}
\wt{g} \ = \ \e^{-\Phi} g \; , \ls
\wt{J} \ = \ \e^{-\Phi} J
\end{equation}
together define a K\"{a}hler manifold. In particular, the Einstein metric
in such special cases can be written as
\begin{equation}
G^{\text{E}}_{MN} \, \d x^M \d x^N
\ = \
\e^{-\Phi/2} \, \eta_{\mu \nu} \, \d x^{\mu} \d x^{\nu}
+ \e^{\Phi/2} \, \wt{g}_{mn} \, \d y^m \d y^n
\end{equation}
so the dilaton plays the role of warp factor in a
familiar form as in type IIB story with 3-fluxes.
For the sake of completeness, we also write the
potential $V$ for the squared Dirac operator in this
case,
\begin{equation}
V \ = \
\frac{3}{2} \, \e^{\Phi}
\nabla_m^2 \e^{-\Phi} + \frac{1}{48}
\left(\d H\right)_{mnpq}\Gamma^{mnpq} \; .
\end{equation}
with $H$ carrying no primitive piece.

\section{Summary}

We have explored the torsionful geometry of the supersymmetric
flux compactification of heterotic string theory.
With the aim at understanding how the low energy gauge sector
arises, and also trying to understand the flux compactification
better, we
isolated the zero mode equation of ten-dimensional gaugino
field, and showed that zero modes responsible for gaugino and
charged matter fermions obey  relatively simple elliptic equations.
Along the way, we found that $\d H=0$ limit always implies
a singular internal manifold, and thus allows the four-dimensional
gaugino to exist despite the formally positive internal operator.
With $\d H \neq 0$ comparable to $H^2$, smooth compactification becomes possible
and we argued why this is a generic behavior by showing
that the higher order term $\alpha' R^2$ is comparable to
$H^2$ and cannot be neglected at least for configurations
near supersymmetric compactifications.

As far as counting the matter content of charged fermions is
concerned, a lesson we learned is that the old
counting of ``generations,'' that is, the number of chiral
charged matter fields, cannot be imported to the compactification
with flux. Recall that the renowned formula where the generation
is given by the Euler number divided by 2 \cite{CHSW85},
replies on the ``minimal embedding'' and $\d H=0$. With $H$ flux,
singularities compromise naive index formula, at least until we
know how to classify and handle the singularities.

Despite the singularity of the manifold, the shape of the
zero mode equations when $\d H=0$ is itself  suggestive.
The  primitive part of the torsion supplies a
spin-independent non-negative potential
to the squared zero mode equation, and its consequence to the
local form of the gauge zero modes might be worth pursuing, in
view of how local physics with a hierarchical warp factor
was important in type IIB compactification.

For more general and
non-singular backgrounds $\d H\neq 0$, it remains to
understand how to solve the Bianchi identity and what this,
together with torsion, implies for index densities in general.
Here we took the first step by constructing the ``Hamiltonian''
$\Delta_{\cal Q}$, relevant for the counting of chiral fermions.
We hope to come
back to study of the index densities, in relation with
the anomaly condition, in  near future.\footnote{See appendix
\ref{index-app} for comments on existing computations for index densities
with torsion.}

\vskip 1cm
\centerline{\bf\large Acknowledgement}
\vskip 1cm

\noindent
We are indebted to Seok Kim for many useful discussions.
PY was supported in part by the Science Research Center Program of
the Korea Science and Engineering Foundation through
the Center for Quantum Spacetime(CQUeST) of
Sogang University with grant number R11-2005-021.

\begin{appendix}

\section*{Appendix}


\section{Conventions}

Conventions for indices are as follows:
\begin{align*}
M, N, \dots & \ \ \text{real ten-dimensional coordinate indices,} \\
A, B, \dots & \ \ \text{real ten-dimensional $SO(9,1)$ indices,} \\
\mu, \nu, \dots & \ \ \text{real four-dimensional coordinate indices,} \\
m, n, \dots & \ \ \text{real six-dimensional coordinate indices,} \\
a,b, \dots & \ \ \text{real six-dimensional $SO(6)$ indices.}
\end{align*}
Antisymmetrization of the indices is defined as
\begin{align}
T_{[M_1 M_2 \cdots M_p]}
\ &= \
\frac{1}{p!} \Big( T_{M_1 M_2 \cdots M_p}
- T_{M_2 M_1 \cdots M_p}
\pm \text{permutations} \Big)
\; .
\end{align}
We adopt the following rule about the contraction of tensors:
\begin{align}
|F_p|^2
\ = \
\frac{1}{p!} \,
g^{M_1 N_1} g^{M_2 N_2} \cdots g^{M_p N_p} \,
F_{M_1 \cdots M_p} F_{N_1 \cdots N_p}
\; .
\end{align}
The $p!$ cancels the sum over permutations of the indices, so that
each independent component appears with coefficient 1.

Vielbeins $e_M{}^A$ and their inverses $E_A{}^M$ from the
curved spacetime metric $g_{MN}$ and the tangent space metric $\eta_{AB}$
are such that
\begin{align*}
&&g_{MN} \ &= \ \eta_{AB} \, e_M{}^A \, e_N{}^B \; , &
\eta_{AB} \ &= \ g_{MN} \, E_A{}^M \, E_B{}^N \; , && \\
&&\delta_M^N \ &= \ e_M{}^A \, E_A{}^N
\; , &
\delta_A^B \ &= \ E_A{}^M \, e_M{}^B
\; . &&
\end{align*}
Here are more conventions and identities related to
differential forms on a $D$-dimensional Riemannian manifold
${\cal M}_D$  \cite{Nakahara}:
\begin{gather*}
\omega_p \ = \
\frac{1}{p!} \omega_{M_1 \cdots M_p}
\, \d x^{M_1} \w \cdots \w \d x^{M_p}
\; , \\
* \omega_p \ = \
\frac{\sqrt{|g_D|}}{p! (D - p)!}
\, \ve_{N_{p+1} \cdots N_D}{}^{M_1 \cdots M_p} \, \omega_{M_1 \cdots M_p}
\, \d x^{N_{p+1}} \w \cdots \w \d x^{N_D}
\; , \\
*1 \ = \ \sqrt{|g_D|} \, \d x^1 \w \cdots \w \d x^D
\; , \\
** \omega_p \ = \
(-1)^{p(D-p)} \omega_p
\; , \\
g_D \, \ve^{M_1 \cdots M_p}{}_{N_{p+1} \cdots N_D} \cdot
\ve_{M_1 \cdots M_p}{}^{L_{p+1} \cdots L_D}
\ = \
p! (D-p)! \cdot \delta^{L_{p+1}}_{[N_{p+1}} \cdots \delta^{L_{D}}_{N_{D}]}
\; ,
\end{gather*}
where $\ve_{M_1 \cdots M_D}$ and $\ve^{M_1 \cdots M_D}$ are tensor
densities.

Finally we close with a useful identity among the Dirac matrices,
which is needed for computation of $-\Slash{D}^2$.
\begin{align}
\begin{split}
&\Gamma^{A_1 A_2 \cdots A_p} \Gamma_{B_1 B_2 \cdots B_q}
\\
& \ls \ = \
\sum_{k=0}^{{\rm min}(p,q)}
(-1)^{\half k (2p -k -1)} \frac{p! q!}{(p-k)! (q-k)! k!}
\delta^{[A_1}_{[B_1} \cdots \delta^{A_k}_{B_k} \Gamma_{\phantom{[B]}}^{A_{k+1} \cdots
      A_p]}{}^{\phantom{[A]}}_{B_{k+1} \cdots B_{q}]}
\; .
\end{split} \label{id-gamma-3}
\end{align}
where
\begin{equation}
\Gamma^{A_1\cdots A_p}
\ = \
\frac{1}{p!}\Big( \Gamma^{A_1}\Gamma^{A_2}\cdots \Gamma^{A_p}
- \Gamma^{A_2}\Gamma^{A_1}\cdots \Gamma^{A_p}
\pm \text{permutations} \Big)
\; .
\end{equation}


\section{Supersymmetry and an $SU(3)$-structure}

Here we summarize the supersymmetry variations of fermions with
zero-th order in $\alpha'$ (the higher order corrections are
shown in \cite{BdR89}):
\bsubeq
\begin{align}
\delta \psi_M \ &= \
\big\{ \p_M + \frac{1}{4} \omega_{-M}{}^{AB} \, \Gamma_{AB} \big\} \epsilon
\; , \\
\delta \lambda \ &= \
- \frac{1}{4} \big\{ \Gamma^M \nabla_M \Phi - \frac{1}{6} H_{MNP} \Gamma^{MNP}
\big\} \epsilon
\; , \\
\delta \chi \ &= \
- \frac{1}{4} F_{MN} \Gamma^{MN} \epsilon
\; .
\end{align}
\esubeq

\subsection{Invariant Forms}

In the heterotic supergravity, we assign the chiralities of fermions
with the followings:
\begin{align}
\Gamma_{(10)} \psi_M \ &= \ + \psi_M \; , \ls
\Gamma_{(10)} \chi \ = \ + \chi \; , \ls
\Gamma_{(10)} \lambda \ = \ - \lambda
\; , \ls
\Gamma_{(10)} \epsilon \ = \ + \epsilon
\; .
\end{align}
The ten-dimensional supersymmetry parameter $\epsilon$, which is a
Majorana-Weyl spinor, decomposes into two kinds of Weyl spinors
under $Spin(9,1) \to Spin(3,1) \times SU(4)$
\begin{align}
\epsilon \ &= \ f \cdot \xi_+ \otimes \eta_+ + f^* \cdot \xi_- \otimes \eta_-
\; ,
\end{align}
where the complex conjugates of these two Weyl spinors are assigned
such as $(\xi_+)^*
= \xi_-$ and $(\eta_+)^* = \eta_-$, respectively; $f$ and $f^*$ are
complex scale factors depending on coordinates.
In this paper we fix these coefficients to 1.
The Weyl spinors on the six-manifold ${\cal M}_6$ define an invariant two-form $J$
and an invariant three-form $\Omega$:
\begin{gather*}
\nabla^{(-)}_m \eta_{\pm} \ = \ 0
\; , \ls
\nabla^{(-)}_m J_{ab} \ = \ \nabla^{(-)}_m \Omega_{abc} \ = \ 0
\; , \\
\eta_{\pm}^{\dagger} \eta_{\pm}^{\phantom{\dagger}} \ = \ 1
\; , \ls J_{ab} \ = \ -i \eta_+^{\dagger} \Gamma_{ab} \eta_+^{\phantom{\dagger}}
\; , \ls
\Omega_{abc} \ = \ \eta_+^{\T} \Gamma_{abc} \eta_+^{\phantom{\T}}
\; .
\end{gather*}
Via the Fierz identity on the Weyl spinors in six-dimensional space,
one can identify $J_m{}^n$ with the almost complex structure and
finds that the metric on the six-dimensional space
becomes hermitian with respect to this almost complex structure:
\begin{align}
J_m{}^p \, J_p{}^n \ &= \ - \delta_m{}^n
\; , \ls
J_m{}^p \, J_n{}^q \, g_{pq} \ = \ g_{mn} \; .
\end{align}
Since there are no invariant five-forms
and there should be one volume form on the $SU(3)$-structure
manifold, these invariant forms satisfy the following equations
\begin{align}
J \w \Omega \ &= \ 0 \; , \ls
J \w J \w J \ = \ \frac{3i}{4} \Omega \w \ol{\Omega}
\ = \
3 ! \sqrt{|g|} \, \d y^1 \w \cdots \w \d y^6
\; .
\end{align}
This $\Omega$ is not a holomorphic three-form, however. See
next subsection.

\subsection{Geometry of Supersymmetric Compactifications}

Supersymmetry variations on the six-manifold ${\cal M}_6$ restrict the
geometrical conditions via relations among the fields
$\{ \Phi, H, F\}$ and the geometrical quantities $\{J, \Omega\}$.
The most typical conditions are given by
\bsubeq
\begin{align}
0 \ &= \
R^{ab}{}_{mn} (\omega_-) J_{ab}
&& (\text{from $\delta \psi_m = 0$})
\; , \label{const-3-2} \\
J_{[m}{}^q \nabla_{|q|} J_{np]}
\ &= \
-2 J_{[m}{}^q J_{n}{}^r H_{p]qr}
&& (\text{from $\delta \psi_m = 0$})
\; , \label{const-3-3} \\
N_{mnp} \ &= \
H_{mnp} - 3 J_{[m}{}^q J_{n}{}^r H_{p]qr}
&& (\text{from $\delta \psi_m = 0$})
\; , \label{const-3-4} \\
N_{mnp} \ &= \ 0
&& (\text{from $\delta \lambda = 0$})
\; , \label{const-3-5} \\
H_{mnp} J^{np} \ &= \
2 J_m{}^q \nabla_q \Phi
&& (\text{from $\delta \lambda = 0$})
\; , \label{const-3-6} \\
0 \ &= \
F_{mn} J^{mn}
&& (\text{from $\delta \chi = 0$})
\; . \label{const-3-7}
\end{align}
\esubeq
By using these we further obtain various simple conditions among the
fields and geometrical quantities in terms of the differential forms \cite{CCDL0306}:
\bsubeq
\begin{align}
0 \ &= \ -2 \d \Phi + \theta
\; , \label{const-4-3} \\
H \ &= \
T^{(\text{B})} \ = \
- \half * \e^{+2 \Phi} \d (\e^{-2 \Phi} J)
\; , \label{const-4-4} \\
0 \ &= \ \d (\e^{-2 \Phi} * J) \ = \ \half \d (\e^{-2 \Phi} J \w J)
\; , \label{const-4-5} \\
0 \ &= \
\d (\e^{-2 \Phi} \Omega)
\; , \label{const-4-6}
\end{align}
which implies that there is a holomorphic three-form
\begin{equation}
\ol{\partial}\left(\e^{-2\Phi}\Omega\right) \ = \ 0 \;.
\end{equation}
\esubeq
Some of quantities above are well-known mathematical objects
for complex geometry. In addition to the familiar Nijenhuis
tensor $N_{mn}{}^p$, the Lee-form $\theta$  and the
Bismut torsion $T^{(\text{B})}_{mnp}$ \cite{Bismut89} are
defined as
\bsubeq
\begin{align}
\theta \ &\equiv \
J \, \lrcorner \, \d J \ = \
\frac{3}{2} J^{mn} \nabla_{[m} J_{np]} \, \d y^p
\; , \label{const-0-1} \\
N_{mn}{}^p \ &\equiv \
J_m{}^q \nabla_{[q} J_{n]}{}^p - J_n{}^q \nabla_{[q} J_{m]}{}^p
\; , \label{const-0-2} \\
T^{(\text{B})}_{mnp} \ &\equiv \
\frac{3}{2} J_m{}^q J_n{}^r J_p{}^s \nabla_{[s} J_{qr]}
\ = \ - \frac{3}{2} J_{[m}{}^q \nabla_{|q|} J_{np]}
\; . \label{const-0-3}
\end{align}
\esubeq
A useful identity for $\d H$ can be found as follows.
Let us decompose $H$-flux on the internal space into the primitive part
$H^0$ and the non-primitive part like
\begin{align}
H \ &= \ H^0 + \frac{1}{4} J \w K \; , \ls
J \, \lrcorner \, H^0 \ = \ J \w H^0 \ = \ 0
\; ,
\end{align}
where $K_m \equiv H_{mnp} J^{np} = 2 J_m{}^n \nabla_n \Phi$ given by
the supersymmetry variation (\ref{const-3-6}).
By using the equations (\ref{const-4-4}), (\ref{const-4-5}) and $J \w H^0 = 0$,
we evaluate the followings:
\bsubeq
\begin{align}
J \w \d H \ &= \
\e^{2 \Phi} \d (\e^{-2 \Phi} J \w H) - \e^{2 \Phi} \d (\e^{-2 \Phi} J) \w H
\; , \\
\e^{2 \Phi} (\d \e^{-2 \Phi} J) \w H
\ &= \ - 2 H \w * H
\ = \
(* 1) \, \frac{1}{3} H_{mnp} H^{mnp}
\; , \\
\begin{split}
\e^{2 \Phi} \d (\e^{-2 \Phi} J \w H)
\ &= \
\frac{1}{4} \e^{2 \Phi} \d (\e^{- 2 \Phi} J \w J \w K)
\\
\ &= \
\half \d K \w * J
\ = \
- * \Big(
\nabla_m^2 \Phi - 2 (\nabla_m \Phi)^2 \Big)
\; .
\end{split}
\end{align}
\esubeq
Thus we obtain an equation among the invariant two-form $J$, the
$H$-flux and the dilaton $\Phi$ such as
\begin{align}
* (J \w \d H)
\ &= \
- \nabla_m^2 \Phi + 2 (\nabla_m \Phi)^2 - \frac{1}{3} H_{mnp} H^{mnp}
\; . \label{H-dilaton}
\end{align}

\subsection{Ricci Scalar Curvature} \label{RicciCurv}

Here we summarize the computation that gives (\ref{Rscalar}).
Starting with
\begin{align}
0 \ &= \ R^{ab}{}_{mn} (\omega_-) J^{mn}
\; ,
\end{align}
where $\omega_- = \omega- H$, so that
\begin{equation}
R_{pmqn} (\omega_-)
\ = \
R_{pmqn} (\omega) - \nabla_q H_{pmn} + \nabla_n H_{pmq}
+ H_{prq} H^{r}{}_{mn} - H_{prn} H^{r}{}_{mq}
\; .
\end{equation}
Contracting with the complex structure one more time,
\begin{align}
0 \ &= \
 R_{pqmn} (\omega_-) J^{pq} J^{mn}
\ = \
3 R_{p[qmn]} (\omega_-) J^{pq} J^{mn}
+ 2 R_{pmqn} J^{pq} J^{mn}
\; ,
\end{align}
The first piece is purely a torsion
\begin{align}
\begin{split}
3 R_{p[qmn]} (\omega_-) J^{pq} J^{mn}
\ &= \
6 (- \nabla_{[m} H_{|p| qn]} + H_{pr[m} H^r{}_{qn]}) J^{pq} J^{mn}
\\
\ &= \
- 6 J^{pq} J^{mn} \nabla_m H_{pqn}
+ 2 J^{pq} J^{mn} H_{pqr} H_{mn}{}^r
\\
\ & \ \ \ \
+ 4 J^{pq} J^{mn} H_{prm} H^r{}_{qn}
\; ,
\end{split}
\end{align}
while the second is the Ricci scalar with torsion
\begin{align}
\begin{split}
2 R_{pmqn} (\omega_-) J^{pq} J^{mn}
\ &= \
2 R_{pmqn} (\omega_-) g^{pq} g^{mn}
\\
\ &= \
2 \left(R(\omega) - H_{mnp} H^{mnp}\right)
\; .
\end{split}
\end{align}
Relations between $J$, $H$ and the dilaton can be used to show
\bsubeq
\begin{align}
J^{pq} J^{mn} H_{prm} H^r{}_{qn}
\ &= \
- \frac{1}{3} H_{mnp} H^{mnp}
\; , \\
J^{pq} J^{mn} H_{pqr} H_{mn}{}^r
\ &= \
4 (\nabla_m \Phi)^2
\; , \\
J^{pq} J^{mn} \nabla_m H_{pqn}
\ &= \
- 2 \nabla_m^2 \Phi - \frac{2}{3} H_{mnp} H^{mnp}
+ 4 (\nabla_m \Phi)^2
\; .
\end{align}
\esubeq
{}Combining these, we find
\begin{align}
\begin{split}
\half R_{pqmn} (\omega_-) J^{pq} J^{mn}
\ &= \
- 3 \left( - 2 \nabla_m^2 \Phi
- \frac{2}{3} H_{mnp} H^{mnp} + 4 (\nabla_m \Phi)^2 \right)
\\
\ & \ \ \ \
+ 4 (\nabla_m \Phi)^2
- \frac{2}{3} H_{mnp} H^{mnp}
+ \big( R (\omega) - H_{mnp} H^{mnp} \big)
\; .
\end{split}
\end{align}
Thus, the supersymmetry demands the scalar curvature
of the internal manifold to satisfy
\begin{align}
 0 \ = \ R (\omega)
+ \frac{1}{3} H_{mnp} H^{mnp}
+ 6 \nabla_m^2 \Phi
- 8 (\nabla_m \Phi)^2
\; .
\end{align}


\section{Equations of Motion}

Equations of motion for $\Phi$, $G_{MN}$, $B_{MN}$ and $\chi$ in
string frame are given as follows:
\bsubeq
\begin{align}
0 \, &= \,
- R (\omega) + \frac{1}{3} H_{MNP} H^{MNP} + 4 (\nabla_M \Phi)^2
- 4 \nabla_M^2 \Phi - {\cal Z}
\; , \\
0 \, &= \,
R_{MN} (\omega) - H_{MPQ} H_N{}^{PQ} + 2 \nabla_M \nabla_N \Phi
\nonumber \\
\ & \ \ \ \,
- \half G_{MN} \Big[ R (\omega) - \frac{1}{3} H_{PQR} H^{PQR}
- 4 (\nabla_P \Phi)^2 + 4 \nabla_P^2 \Phi + {\cal Z} \Big]
\nonumber \\
\ & \ \ \ \,
- 2 \alpha' \Big[
\tr (F_{MP} F_N{}^P) - \tr \{ R_{MP} (\omega_+) R_N{}^P (\omega_+) \}
\Big]
\nonumber \\
\ & \ \ \ \,
- 2 \alpha' \e^{2 \Phi} \Big[
2 \nabla^P \nabla^Q_{(+)} \big\{ \e^{- 2 \Phi} R_{MPNQ} (\omega_+) \big\}
- \nabla^Q_{(+)} \big\{ \e^{-2 \Phi} R_{MPQR} (\omega_+) \big\} H_N{}^{PR}
\nonumber \\
\ & \LS \ls \ \ \,
- 2 \nabla^P \big\{ \e^{- 2 \Phi} R_{MPQR} (\omega_+) H_N{}^{QR} \big\}
- 2 \e^{-2 \Phi} R_{MPQR} (\omega_+) H_N{}^{PS} H_{S}{}^{QR}
\nonumber \\
\ & \LS \ls \ \ \,
- \nabla^P \nabla^Q_{(+)} \big\{ \e^{- 2 \Phi} R_{MNPQ} (\omega_+) \big\}
+ \nabla^P \big\{ \e^{-2 \Phi} R_{MNQR} (\omega_+) H_P{}^{QR} \big\}
\Big]
\; , \\
0 \, &= \,
\nabla^M (\e^{-2 \Phi} H_{MNP})
\; , \\
0 \, &= \,
\Slash{D} (\omega,A) \chi
- \frac{1}{12} \Gamma^{MNP} \chi H_{MNP}
\nonumber \\
 & \ \ \ \,
- \Gamma^M \chi \nabla_M \Phi
+ \frac{3}{2} \Gamma^M \Gamma^{NP} ( F_{NP} + \hat{F}_{NP} )
\Big( \psi_M + \frac{2}{3} \Gamma_M \lambda \Big)
\; , \\
\text{where} &\ls {\cal Z} \, \equiv \,
- \alpha' \Big[ \tr (F_{MN} F^{MN}) - \tr \{ R_{MN} (\omega_+) R^{MN}
  (\omega_+) \} \Big]
\; .
\end{align}
\esubeq
Notice that we defined the trace with respect to the former two indices of
the curvature tensors such as
$\tr \{ R_{MN} (\omega_+) R^{MN} (\omega_+) \} =
- R_{PQMN} (\omega_+) R^{PQMN} (\omega_+)$.
Via the anomaly cancellation
in ten dimensions, the Bianchi identity of $H$-flux is given by (see
\cite{Hull86, BdR89})
\begin{align}
\d H \ &= \ \alpha' \Big[
\tr \{ R (\omega_+) \w R (\omega_+) \} - \tr (F \w F)
\Big]
\; .
\label{Bianchi-H}
\end{align}

\section{Index Densities with Torsion} \label{index-app}

It is often stated that introduction of torsion does
not affect index. This is natural since the torsion
piece, as far as the classical geometry goes, can be
thought of a continuous deformation on the Dirac operator,
under which an index of Fredholm operator should be
invariant.

We should not be mislead to expect from this mathematical
statement that
flux has no effect in the fermion counting in string
compactification. Fluxes in string compactification
can affect the fermion counting in two qualitative
ways. One is to modify the Dirac equation so that
fermions of different kind (or chirality) get mixed up and
usual chirality operator cannot be used
to define an index. Another, which is relevant for
the gauge sector fermions in our heterotic theory,
is the fact that the geometry can backreact to the flux
in some essential way. This was the case for flux
compactification with $\d H=0$, as we saw above.

With these said, it is still curious that index density
formula for an arbitrary smooth manifold with torsion
seem not available. The closest work to this can be
found in \cite{Mav88}, which computes the
Atiyah-Singer index densities when the manifold has
a completely anti-symmetric torsion which is closed.
According to this work, the Atiyah-Singer index density
for $\Slash{D}(\hat{\omega}, A)$ with $\d H=0$
would be given by the usual characteristic polynomial \cite{Mav88, Yajima89}
\begin{equation}
{\cal A}(R_+) \wedge ch(F)
\end{equation}
with A-genus ${\cal A}$ and the Chern character $ch$ of the
gauge bundle. Note that in place of the curvature two-form, we
have the curvature two-form $R_+$ of the connection $\omega_+
=\omega+H$, rather than $R_-$.

Assuming that the path integral approach
taken there is accurate, one may understand switching as follows.
Recall that the index density formula is obtained by using
the identity
\begin{equation}
\text{index} \Slash{D}
\ \equiv \
\lim_{\beta \to \infty}
\left(\e^{-\beta \Delta} \prod_{a=1}^6 \Gamma_a \right)
\ = \
\lim_{\beta \to 0} \left(\e^{-\beta \Delta}\prod_{a=1}^6 \Gamma_a \right)
\; ,
\end{equation}
which holds provided that the spectrum of $\Slash{D}$ is
discrete. Realizing $\Delta$ as a Hamiltonian of a supersymmetric
quantum mechanics with supercharge $\Slash{D}$, one obtains the
metric-dependent part of the index density from one-loop
determinant of bosonic oscillation. The curvature part of the
index densities are built with monomials like
\begin{equation}
\tr \left\{ R_{abm}{}^n (\omega_-) \, e^a \wedge e^b \right\}^k
\end{equation}
with vielbein one-forms $e^a$.
Here the trace is taken over $m$ and $n$ indices, the coordinate
indices. Note that this is opposite of usual invariant density
where the Lie algebra indices $a$ and $b$ are traced
over. Without
torsion, this flip does not matter since the two sets of indices are
interchangeable. With torsion, instead, we have
\begin{align}
R_{mnpq} (\omega_+)
\ = \ R_{pqmn} (\omega_-) + (\d H)_{pqmn}
\ = \ R_{pqmn} (\omega_-)
\; .
\end{align}
Thus, the formal computation yields an invariant polynomial of type
\begin{equation}
\tr \left\{ R(\omega_+=\omega+H)\right\}^{2k}
\end{equation}
provided that $\d H=0$.

If the manifold were smooth and compact, this
would have demonstrated that the index is independent of $H$.
To see this, let us define an $SO$-valued torsion one-form as
\begin{equation}
T^a{}_{b} \ = \ H_m{}^{a}{}_{b} \, \d y^m
\; .
\end{equation}
It can be seen easily that
\begin{equation}
\tr \left\{ R(\omega_+) \right\}^{2k}
\ = \
\tr \{ R(\omega) \}^{2k} + \d \left(2k \int_0^1 \! \d x \,
\tr \left\{ T \wedge R(\omega + xT)^{2k-1} \right\} \right)
\; ,
\end{equation}
which shows that the torsion contribution will integrate to
zero on a compact and smooth manifold.

With $\d H\neq 0$, the quantum mechanics is somewhat modified
because of the quartic terms that survives in $\Delta$, in the
form $\sim (\d H)_{abcd}\Gamma^{abcd}$ \cite{NY81, CZ97}. Naive extension of
Mavromatos' computation is not difficult to carry out, but it is not clear whether
the final formula makes sense. $\d H$ can enter in two distinct
ways: First is a further shift of the curvature tensor $R (\omega_+)$ to
$R (\omega_+) - \d H$. Note that this is because that curvature is
actually $R (\omega_-)$ with the coordinate
indices and the Lie-algebra indices flipped. Also $\d H$
makes appearance as a factor of $\d H/\beta$ outside the trace,
where all 4 indices should be regarded as coordinate indices.
Contribution like the latter must disappear upon integration on the six-manifold,
yet explicit check of this has not been performed. In fact,
it is not clear if the resulting formula following this line
of derivation makes any sense as a
topological density.

One reason for such difficulties must be due to the subtlety
in the regularization of the path integral approach. Somewhat
formal manipulation, originally due to \cite{AG83,AG83'},
seems to
fail for manifold with torsion. In literature, rigorous
computations of this kind exists only in the context of
four-dimensional spacetime, largely in connection with
axial anomaly in quantum field theory. See \cite{PW99}
for detailed and rigorous computation that demonstrates
that torsion contributes a total derivative term only in
four-dimensional case. For rigorous
computation of index densities, a heat-kernel approach \cite{Obukhov83}
would be more desirable which is not yet available for
six and higher dimensions.

\end{appendix}


}

\begin{thebibliography}{99}

\bibitem{Susskind}
  L.~Susskind,
  ``{\sl The Anthropic Landscape of String Theory},''
{\tt hep-th/0302219}.

  \bibitem{Douglas}
  M.~R.~Douglas,
  ``{\sl The Statistics of String/M-theory Vacua},''
  JHEP {\bf 0305} (2003) 046,
{\tt hep-th/0303194}.


\bibitem{Ashok}
  S.~Ashok and M.~R.~Douglas,
  ``{\sl Counting Flux Vacua},''
  JHEP {\bf 0401} (2004) 060,
  {\tt hep-th/0307049}.


\bibitem{Denef}
  F.~Denef and M.~R.~Douglas,
  ``{\sl Distributions of Flux Vacua},''
  JHEP {\bf 0405} (2004) 072,
  {\tt hep-th/0404116}.


\bibitem{DRS99}
  K.~Dasgupta, G.~Rajesh and S.~Sethi,
  ``{\sl M-theory, Orientifolds and $G$-flux},''
  JHEP {\bf 9908} (1999) 023,
  {\tt hep-th/9908088}.

\bibitem{KST02}
  S.~Kachru, M.~B.~Schulz and S.~Trivedi,
  ``{\sl Moduli Stabilization from Fluxes in a Simple IIB Orientifold},''
  JHEP {\bf 0310} (2003) 007,
  {\tt hep-th/0201028}.

\bibitem{BB96}
K.~Becker and M.~Becker,
  ``{\sl M-theory on Eight-manifolds},''
  Nucl.\ Phys.\ B {\bf 477} (1996) 155,
  {\tt hep-th/9605053}.

\bibitem{GKP0105}
  S.~B.~Giddings, S.~Kachru and J.~Polchinski,
  ``{\sl Hierarchies from Fluxes in String Compactifications},''
  Phys.\ Rev.\ D {\bf 66} (2002) 106006,
  {\tt hep-th/0105097}.

\bibitem{KS00}
I.~R.~Klebanov and M.~J.~Strassler,
  ``{\sl Supergravity and a Confining Gauge Theory: Duality Cascades and
  $\chi$SB-resolution of Naked Singularities},''
  JHEP {\bf 0008} (2000) 052,
  {\tt hep-th/0007191}.

\bibitem{CHSW85}
  P.~Candelas, G.~T.~Horowitz, A.~Strominger and E.~Witten,
  ``{\sl Vacuum Configurations for Superstrings},''
  Nucl.\ Phys.\ B {\bf 258} (1985) 46.

\bibitem{RS9905}
L.~Randall and R.~Sundrum,
  ``{\sl A Large Mass Hierarchy from a Small Extra Dimension},''
  Phys.\ Rev.\ Lett.\  {\bf 83} (1999) 3370,
  {\tt hep-ph/9905221}.

\bibitem{KKLT}
  S.~Kachru, R.~Kallosh, A.~Linde and S.~P.~Trivedi,
  ``{\sl De Sitter Vacua in String Theory},''
  Phys.\ Rev.\ D {\bf 68}(2003) 046005,
  {\tt hep-th/0301240}.

\bibitem{CCDLMZ0211}
  G.~L.~Cardoso, G.~Curio, G.~Dall'Agata, D.~L\"{u}st, P.~Manousselis and G.~Zoupanos,
  ``{\sl Non-K\"{a}hler String Backgrounds and Their Five Torsion Classes},''
  Nucl.\ Phys.\ B {\bf 652} (2003) 5,
  {\tt hep-th/0211118}.

\bibitem{Hit0209}
N.~J.~Hitchin,
``{\sl Generalized Calabi-Yau Manifolds},''
{\tt math.DG/0209099}.

\bibitem{GMW0302}
  J.~P.~Gauntlett, D.~Martelli and D.~Waldram,
  ``{\sl Superstrings with Intrinsic Torsion},''
  Phys.\ Rev.\ D {\bf 69} (2004) 086002,
  {\tt hep-th/0302158}.


\bibitem{FMT}
  S.~Fidanza, R.~Minasian and A.~Tomasiello,
  ``{\sl Mirror Symmetric $SU(3)$-structure Manifolds with NS Fluxes},''
  Commun.\ Math.\ Phys.\  {\bf 254} (2005) 401,
  {\tt hep-th/0311122}.


\bibitem{Gua0401}
M.~Gualtieri,
``{\sl Generalized Complex Geometry},''
{\tt math.DG/0401221}.


\bibitem{GMPT}
  M.~Gra\~{n}a, R.~Minasian, M.~Petrini and A.~Tomasiello,
  ``{\sl Supersymmetric Backgrounds from Generalized Calabi-Yau Manifolds},''
  JHEP {\bf 0408} (2004) 046,
  {\tt hep-th/0406137}.



\bibitem{GMPT2}
  M.~Gra\~{n}a, R.~Minasian, M.~Petrini and A.~Tomasiello,
  ``{\sl Generalized Structures of ${\cal N} = 1$ vacua},''
  JHEP {\bf 0511} (2005) 020,
  {\tt hep-th/0505212}.

\bibitem{Grana0509}
  M.~Gra\~{n}a,
  ``{\sl Flux Compactifications in String Theory: A Comprehensive Review},''
  Phys.\ Rept.\  {\bf 423} (2006) 91,
  {\tt hep-th/0509003}.

\bibitem{MN0007}
  J.~M.~Maldacena and C.~Nu\~{n}ez,
  ``{\sl Supergravity Description of Field Theories on Curved
  Manifolds and a No Go Theorem},''
  Int.\ J.\ Mod.\ Phys.\ A {\bf 16} (2001) 822,
  {\tt hep-th/0007018}.



\bibitem{GS84}
M.~B.~Green and J.~H.~Schwarz,
  ``{\sl Anomaly Cancellation in Supersymmetric $D=10$ Gauge Theory and Superstring
  Theory},''
  Phys.\ Lett.\ B {\bf 149} (1984) 117.

\bibitem{BdR89}
E.~A.~Bergshoeff and M.~de Roo,
``{\sl The Quartic Effective Action of the Heterotic String and Supersymmetry},''
  Nucl.\ Phys.\ B {\bf 328} (1989) 439.

\bibitem{Str86}
  A.~Strominger,
  ``{\sl Superstrings with Torsion},''
Nucl.\ Phys.\  B {\bf 274} (1986) 253.

\bibitem{BBGDS}
  K.~Becker, M.~Becker, P.~S.~Green, K.~Dasgupta and E.~Sharpe,
  ``{\sl Compactifications of Heterotic Strings on Non-K\"{a}hler
  Complex  Manifolds. II},''
  Nucl.\ Phys.\ B {\bf 678} (2004) 19,
  {\tt hep-th/0310058}.

\bibitem{BBDS}
  K.~Becker, M.~Becker, K.~Dasgupta and P.~S.~Green,
  ``{\sl Compactifications of Heterotic Theory on non-K\"{a}hler Complex Manifolds. I},''
  JHEP {\bf 0304} (2003) 007,
  {\tt hep-th/0301161}.

\bibitem{FY0604}
  J.~X.~Fu and S.~T.~Yau,
  ``{\sl The Theory of Superstring with Flux on Non-K\"{a}hler Manifolds and the
  Complex Monge-Amp\`{e}re Equation},''
  {\tt hep-th/0604063}.

\bibitem{BBFTY0604}
  K.~Becker, M.~Becker, J.~X.~Fu, L.~S.~Tseng and S.~T.~Yau,
  ``{\sl Anomaly Cancellation and Smooth Non-K\"{a}hler Solutions in Heterotic String
  Theory},''
  {\tt hep-th/0604137}.


\bibitem{Dine}
  M.~Dine, R.~Rohm, N.~Seiberg and E.~Witten,
  ``{\sl Gluino Condensation in Superstring Models},''
  Phys.\ Lett.\ B {\bf 156} (1985) 55.


\bibitem{Lop}
  G.~Lopes Cardoso, G.~Curio, G.~Dall'Agata and D.~L\"{u}st,
  ``{\sl Heterotic String Theory on Non-K\"{a}hler Manifolds with $H$-flux and Gaugino
  Condensate},''
  Fortsch.\ Phys.\  {\bf 52} (2004) 483,
  {\tt hep-th/0310021}.

\bibitem{FL0507}
  A.~R.~Frey and M.~Lippert,
  ``{\sl AdS Strings with Torsion: Non-complex Heterotic Compactifications},''
  Phys.\ Rev.\ D {\bf 72} (2005) 126001,
  {\tt hep-th/0507202}.

\bibitem{MPZ0511}
  P.~Manousselis, N.~Prezas and G.~Zoupanos,
  ``{\sl Supersymmetric Compactifications of Heterotic Strings with Fluxes and
  Condensates},''
Nucl.\ Phys.\ B {\bf 739} (2006) 85,
  {\tt hep-th/0511122}.

\bibitem{CCDL0306}
  G.~L.~Cardoso, G.~Curio, G.~Dall'Agata and D.~L\"{u}st,
  ``{\sl BPS Action and Superpotential for Heterotic String Compactifications  with
  Fluxes},''
  JHEP {\bf 0310} (2003) 004,
  {\tt hep-th/0306088}.

\bibitem{Bismut89}
J.~M.~Bismut,
``{\sl A Local Index Theorem for Non-K\"{a}hler Manifolds},''
Math.\ Ann.\ {\bf 284} (1989) 681.

\bibitem{GVW99}
S.~Gukov, C.~Vafa and E.~Witten,
  ``{\sl CFT's from Calabi-Yau Four-folds},''
  Nucl.\ Phys.\ B {\bf 584} (2000) 69
  [Erratum-ibid.\ B {\bf 608} (2001) 477],
  {\tt hep-th/9906070}.

\bibitem{Mav88}
  N.~E.~Mavromatos,
  ``{\sl A Note on the Atiyah-Singer Index Theorem for Manifolds with Totally
  Antisymmetric $H$ Torsion},''
J.\ Phys.\ A {\bf 21} (1988) 2279.


\bibitem{Hull86}
C.~M.~Hull,
``{\sl Compactifications of the Heterotic Superstring},''
Phys.\ Lett.\ B {\bf 178} (1986) 357.

\bibitem{Ivanov}
  S.~Ivanov and G.~Papadopoulos,
  ``{\it A No-go Theorem for String Warped Compactifications},''
  Phys.\ Lett.\ B {\bf 497} (2001) 309,
  {\tt hep-th/0008232}.


  \bibitem{Gillard}
  J.~Gillard, G.~Papadopoulos and D.~Tsimpis,
  ``{\sl Anomaly, Fluxes and $(2,0)$ Heterotic-string Compactifications},''
  JHEP {\bf 0306} (2003) 035,
  {\tt hep-th/0304126}.

\bibitem{Gran}
  U.~Gran, P.~Lohrmann and G.~Papadopoulos,
  ``{\sl The Spinorial Geometry of Supersymmetric Heterotic String Backgrounds},''
  JHEP {\bf 0602} (2006) 063,
  {\tt hep-th/0510176}.

\bibitem{DW}
  O.~DeWolfe and S.~B.~Giddings,
  ``{\sl Scales and Hierarchies in Warped Compactifications and Brane Worlds},''
  Phys.\ Rev.\ D {\bf 67} (2003) 066008,
  {\tt hep-th/0208123}.

\bibitem{BT0509}
  K.~Becker and L.~S.~Tseng,
  ``{\sl Heterotic Flux Compactifications and Their Moduli},''
Nucl.\ Phys.\ B {\bf 741} (2006) 162,
  {\tt hep-th/0509131}.

\bibitem{Nakahara}
M.~Nakahara,
``{\sl Geometry, Topology and Physics},''
Institute of Physics Publishing (1990), Bristol.


\bibitem{Yajima89}
S.~Yajima,
  ``{\sl Gravitational Anomalies with Curl Vanishing Torsion},''
Hiroshima University Preprint {\tt HUPD-8901} (1989).

\bibitem{NY81}
  H.~T.~Nieh and M.~L.~Yan,
  ``{\sl An Identity in Riemann-Cartan Geometry},''
  J.\ Math.\ Phys.\  {\bf 23} (1982) 373.

\bibitem{CZ97}
  O.~Chandia and J.~Zanelli,
  ``{\sl Topological Invariants, Instantons and Chiral Anomaly on Spaces with
  Torsion},''
  Phys.\ Rev.\ D {\bf 55} (1997) 7580,
  {\tt hep-th/9702025}.

\bibitem{AG83}
  L.~Alvarez-Gaum\'{e},
  ``{\sl Supersymmetry and the Atiyah-Singer Index Theorem},''
  Commun.\ Math.\ Phys.\  {\bf 90} (1983) 161.

\bibitem{AG83'}
  L.~Alvarez-Gaum\'{e},
  ``{\sl A Note on the Atiyah-Singer Index Theorem},''
  J.\ Phys.\ A {\bf 16} (1983) 4177.

\bibitem{PW99}
  K.~Peeters and A.~Waldron,
  ``{\sl Spinors on Manifolds with Boundary: APS Index Theorems with Torsion},''
  JHEP {\bf 9902} (1999) 024,
  {\tt hep-th/9901016}.

\bibitem{Obukhov83}
Y.~N.~Obukhov,
  ``{\sl Spectral Geometry of the Riemann-Cartan Space-Time},''
  Nucl.\ Phys.\ B {\bf 212} (1983) 237.




\end{thebibliography}
\end{document}